\documentclass{moriond}

\def\Journalref#1#2#3#4{{#1}, {#2}, #3, #4}
\def\Arxivref#1#2#3{#1, {#2}, #3}
\def\tbs#1#2{#1 {#2}}

\def\AA{A\&A}
\def\Arx{ArXiv e-prints}

\def\be{\begin{equation}}
\def\ee{\end{equation}}
\def\bea{\begin{eqnarray}}
\def\eea{\end{eqnarray}}

\newcommand{\Photo}{}

\begin{document}
\vspace*{4cm}
\title{Can the CIB constrain the dark energy?}

\author{ A. Maniyar$^1$, G. Lagache$^1$, M. B\'ethermin$^1$, S. Ili\'c$^2$}

\address{$^1$Aix Marseille Univ, CNRS, CNES, LAM, Marseille, France \\ $^2$CEICO, Institute of Physics of the Czech Academy of Sciences, Na Slovance 2, Praha 8 Czech Republic}

\maketitle\abstracts{
Galaxies are often used as tracers of the large scale structure (LSS) to measure the Integrated Sachs-Wolfe effect (ISW) by cross-correlating the galaxy survey maps with the Cosmic Microwave Background (CMB) map. We use the Cosmic Infrared Background (CIB) as a tracer of the LSS to perform a theoretical CIB-CMB cross-correlation to measure the ISW for different Planck HFI frequencies. We discuss the detectability of this ISW signal using a Signal-to-noise ratio analysis and find that the ISW detected this way can provide us with the highest SNR for a single tracer ranging from 5 to 6.7 (maximum being for 857 GHz) with the CIB and CMB maps extracted over the whole sky. A Fisher matrix analysis showed that this measurement of the ISW can improve the constraints on the cosmological parameters; especially the equation of state of the dark energy $w$ by $\sim 47\%$. Performing a more realistic analysis including the galactic dust residuals in the CIB maps over realistic sky fractions shows that the dust power spectra dominate over the CIB power spectra at $\ell < 100$ and ISW can't be detected with high SNR. We perform the cross-correlation on the existing CIB-CMB maps over $\sim 11\%$ of the sky in the southern hemisphere and find that the ISW is not detected with the existing CIB maps over such small sky fractions.}

\section{Integrated Sachs-Wolfe effect and Cosmic Infrared Background}
ISW effect is a potential powerful probe of the dark energy. It's the low-redshift counterpart of the Sachs-Wolfe effect happening at the last scattering surface. 
The ISW occurs in an Universe not dominated by matter. As the dark energy starts dominating, it causes the gravitational potential wells and hills on large scales to decay. As a result, 
the CMB photons travelling across them undergo a net gain or loss of energy for potential wells and hills respectively. It is the dominant contribution to the CMB power spectrum at 
large angular scales and has a very small amplitude. Unfortunately, at these scales, the statistical noise due to cosmic variance is of the same order of magnitude as 
the signal making the ISW detection in the CMB alone extremely challenging. A non-zero spatial correlation between the CMB anisotropies and tracers of the large scale matter distribution is expected, and can be used to extract the ISW signal. In this study, we follow the approach of Ili\'c et al(2014)\cite{steph} and 
use the CIB as a LSS tracer for this purpose. The CIB is the weighted integral of the dust heated by the young UV-bright stars within the 
galaxies through the cosmic time. Anisotropies in the CIB trace the large scale distribution of the galaxy density field and hence, the underlying distribution of the dark matter haloes 
which host these galaxies, up to a bias factor. Therefore, the CIB can be used as a tracer of the LSS. 
\subsection{CIB, CIB-CMB lensing modeling}
The ISW dominates at the lower multipoles (large angular scales) where the clustering is dominated by the correlation between the dark matter haloes and non-linear terms can be 
ignored. We use the linear CIB model presented in Maniyar et al (2018)\cite{man}. The linear power spectrum is:
\begin{equation}
C_l^{\nu \times \nu'} =  \int \frac{dz}{\chi^2} \frac{d\chi}{dz}a^2 b_{eff}^2 \bar{j}(\nu,z) \bar{j}(\nu',z)P_{lin}(k = l/\chi,z)
\label{eq:cib}
\end{equation}
where $P_{lin}(k,z)$ is the linear theory dark matter power spectrum, $\chi(z)$ is the comoving distance to redshift $z$ and $a = 1/(1+z)$ is the scale factor. $b_{eff}$ is the mean bias 
of dark matter haloes hosting dusty galaxies contributing to the CIB at a given redshift weighted by their contribution to the emissivities (Planck Collaboration (2014)\cite{pc} and Maniyar et al (2018)\cite{man}) and is called the 
effective bias. $\bar{j}$ is the mean comoving emissivity of the CIB galaxies. It can be seen from Eq.\,\ref{eq:cib} that $b_{eff}$ and $\bar{j}$ are degenerate.
To break this degeneracy, we utilise the CIB-CMB lensing cross-correlation which is given by
\begin{equation}
C_l^{\nu\phi} = \int b_{eff} \bar{j}(\nu,z) \frac{3}{l^2}\Omega_m H_0^2 \bigg(\frac{\chi_* - \chi}{\chi_*\chi}\bigg) P_{lin}(k = l/\chi,z) d\chi \,,
\label{eq:lens}
\end{equation}
where $\chi_*$ is the comoving distance to the CMB last scattering surface, $\Omega_m$ is the cosmological matter density and $H_0$ is the value of the Hubble's constant today. 
From Eq.\,\ref{eq:lens}, we see that $C_l^{\nu,\phi}$ is proportional to $b_{eff}$, whereas, $C_l^{CIB}$ is proportional to $b_{eff}^2$. Thus, using also the CIB-CMB lensing potential 
correlation in the likelihood helps us partially resolve the degeneracy between $b_{eff}$ and $\bar{j}$ parameters. Details of all the parameters of the model and various data-sets used to fit the model are given in Maniyar et al (2018)\cite{man} and Maniyar et al (in prep)\cite{man2}.
\section{CIB-ISW cross-correlation and cosmological constraints}
The SNR for the ISW obtained through the CIB-CMB cross-correlation is given as:
\begin{equation}\label{eq:snr_real}
\Big[\frac{S}{N}\Big]^2(\nu) = \sum_{\ell = 2}^{\ell_{max}} (2\ell + 1) \frac{f_\mathrm{sky}\,\big[C_\ell^\mathrm{CIBxCMB}(\nu)\big]^2 }{\big[C_\ell^\mathrm{CIBxCMB}(\nu)\big]^2 + \big[C_\ell^{\mathrm{CIB}}(\nu) + N_\ell^{\mathrm{CIB}}(\nu)\big]\, C_\ell^{\mathrm{CMB}}}
\end{equation}
where the N$_\ell^\mathrm{CIB}(\nu)$ term contains the galactic dust residuals left in the CIB maps and f$_\mathrm{sky}$ represents the fraction of the sky common to the CMB and the CIB maps. 
\begin{figure}
\begin{minipage}{0.5\linewidth}
\centerline{\includegraphics[width=\linewidth]{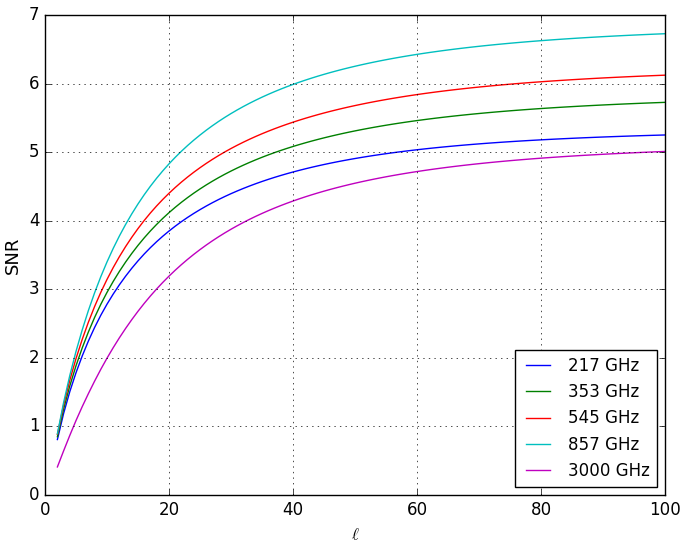}}
\end{minipage}
\hfill
\begin{minipage}{0.5\linewidth}
\centerline{\includegraphics[width=\linewidth]{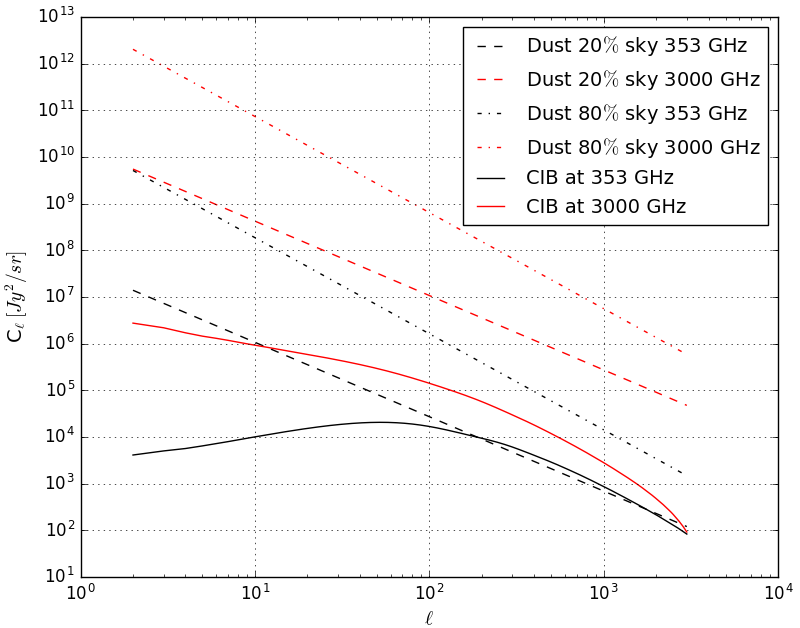}}
\end{minipage}
\caption[]{Left panel shows the predicted cumulative SNR for different Planck frequencies in the ideal case. The right panel shows the power spectra of the CIB and the galactic dust at 353\,GHz and 3000\,GHz with black and red lines respectively. The dust power spectra have been calculated over 20, and 80\% of the sky.}
\label{fig:cross_snr_dust}
\end{figure}
\subsection{Ideal case}
The ideal case assumes that the CIB and the CMB maps are extracted over the full sky and are completely dust-free i.e. N$_\ell^\mathrm{CIB}(\nu) =0$ and f$_\mathrm{sky} =1$. The only limiting factor is the cosmic variance and we get the highest
SNR in this scenario. \\
Left panel of Fig.\,\ref{fig:cross_snr_dust} shows the cumulative SNR for the ISW at each frequency. We observe that the multipoles up to 
$\ell \leq 50$ contribute the most to the SNR and it could reach as high as 6.7 for 857 GHz. With a signal observed at such high SNR, the CIB-ISW cross-correlation can 
be a competitive probe of the cosmology and the dark energy. We therefore perform a joint Fisher matrix analysis with the CMB, the CIB-CMB lensing and CIB-ISW cross-correlation to 
compute the improvement in the constraints on the cosmological parameters. Combining the information from the CIB and CMB alone in a Fisher matrix provides with constraints on the cosmological parameters. Adding the information from the CIB-CMB lensing cross-correlation as well as the ISW obtained through the CIB-CMB cross-correlation further improves these constraints. 
\begin{table}[t]
\caption{The predicted improvement in the constraints on the cosmological parameters after adding the CIB$\times$Lensing and the ISW signal in the CIB + CMB Fisher matrix for 100\% sky for $\Lambda$CDM model.}
\vspace{0.4cm}
\label{tab:lcdm}
\begin{center}
\begin{tabular}{|c|c|c|c|c|c|c|}
\hline
\multicolumn{7}{|c|}{\% improvement on the CIB+CMB constraints with the ISW over 100\% sky}\\[2pt]
\hline 
Parameters & H$_0$ & $\Omega_bh^2$ & $\Omega_ch^2$ & $\tau$ & n$_s$ & 10$^9$A$_s$\\[2pt]
[CIB + CMB] + CIB$\times$Lensing & 1.62 & 0.79 & 1.49 & 4.58 & 0.58 & 5.12 \\[2pt]
[CIB + CMB + CIB$\times$Lensing] + ISW & 6.86 & 2.32 & 8.77 & 16.9 & 4.77 & 16.29 \\[2pt]
\hline
\end{tabular}
\end{center}
\end{table}
\begin{table}[t]
\caption{The predicted improvement in the constraints on the cosmological parameters after adding the CIB$\times$Lensing and the ISW signal to the CIB+CMB+BAO+SNe Ia+H$_0$ prior for 100\% sky for the $w$CDM model.}
\label{tab:wcdm}
\vspace{0.1cm}
\begin{center}
\begin{tabular}{|c|c|c|c|c|c|c|c|}
\hline 
\multicolumn{8}{|c|}{\% improvement on the CIB+CMB+Ext constraints with the ISW over 100\% sky}\\[2pt]
\hline 
 Parameters & H$_0$ & $w$ & $\Omega_bh^2$ & $\Omega_ch^2$ & $\tau$ & n$_s$ & 10$^9$A$_s$\\[2pt]
[CIB + CMB + Ext] + CIB$\times$Lensing & 3.83 & 2.60 & 0.33 & 0.52 & 3.67 & 0.24 & 4.20 \\[2pt]
[CIB + CMB + Ext + CIB$\times$Lensing] + ISW & 31.06 & 49.17 & 5.13 & 26.35 & 22.64 & 12.41 & 21.11 \\[2pt]
\hline
\end{tabular}
\end{center}
\end{table} 
Table\,\ref{tab:lcdm} shows this improvement in the context of the $\Lambda$CDM model. We observe an improvement of 6\% on the constraints on $\Omega_m$ (and hence $\Omega_\Lambda$) and of 11\% on $\tau$ (and hence $10^9A_s$ as they are positively correlated) after adding the ISW information alone on top of the CIB + CMB + CIB-CMB lensing cross-correlation information.  \\
In a flat Universe, the ISW directly results from the dark energy and it has a strong dependence on the equation of state of the dark energy $w = P/\rho$. In the $\Lambda$CDM model, $w$ is 
kept constant at -1. In the $w$CDM model, $w$ is a free parameter along with the 6 other cosmological parameters of the $\Lambda$CDM model. Due to the geometrical degeneracy between $w$ and H$_0$, CMB data alone can not constrain both these parameters together. Therefore, external data sets like baryonic acoustic oscillations (BAO) and supernovae type Ia (SNe Ia), with the low redshift distance measurements are used with the CMB data to break this degeneracy. As in the case of the $\Lambda$CDM 
model, Tab.\,\ref{tab:wcdm} shows the improvement in the constraints on the cosmological parameters in the context of the $w$CDM model. We observe an improvement of 47\% on the constraints on $w$ and of 18\% on $\tau$ and $10^9A_s$ after adding the ISW information alone on top of the CIB + CMB + CIB-CMB lensing cross-correlation + Ext 
(BAO + SNE Ia + H$_0$ prior) data.
\subsection{Realistic case}
On a large part of the sky, we have strong emission from our own Galaxy. This emission is much higher than the CIB signal. This prevents us to extract the CIB signal over the part of the sky where the galactic plane resides and reduces the available sky fraction for measurement by 30\%. Even in the remaining part of the sky, the CIB maps are contaminated by dust that needs to be removed. This removal leaves residual noise and has to be accounted for in the analysis. We calculate the 
dust power spectra at different frequencies for different sky fractions. Right panel of Fig.\,\ref{fig:cross_snr_dust} shows the CIB and the galactic dust power spectra at 353 and 3000 GHz for different sky fractions. The dust power spectra dominate over the CIB power spectra at lower multipoles where we expect to extract the ISW signal. As mentioned in Planck Collaboration (2014)\cite{pc}, the dust residuals are of the order of 5-10\% on the power spectrum. In our analysis, we take the conservative 10\% dust residuals of C$_\ell^\mathrm{dust}(\nu)$ as the noise term. From Eq.\,\ref{eq:snr_real}, we see that the N$_\ell^\mathrm{CIB}(\nu)$ term is quite big compared to the ideal noiseless ideal case and the SNR drops quite drastically with values from 0.5 for 3000 GHz to 1.5 for 353 GHz for 20\% of the sky. 
\section{Conclusions}
Using the CIB as a tracer of the LSS, we calculate the theoretical prediction for the cross-correlation of the CIB with the CMB to extract an ISW signal. We find that the ISW could be obtained with the highest SNR (reaching up to 6.7 for 857 GHz) for a single tracer with this technique when the CIB and CMB maps are dust-free and are extracted over the whole sky. We show that the ISW detected at such high SNR can improve the constraints on the cosmological parameters quite considerably; especially the constraints on the equation of state of the dark energy are improved by $\sim 47\%$. However, in the realistic case, we are limited by the presence of the galactic dust residuals in the maps and also by the available sky fractions over which the maps are extracted. When the same analysis is performed with the addition of 10\% galactic dust residual power spectra, they completely dominate over the CIB at lower multipoles of interest and obtaining an ISW signal with high SNR is not possible anymore. In the end, we also performed the cross-correlation of the CIB and the CMB Planck maps over $\sim 11\%$ of the southern sky in a field named 'GASS' (Planck Collaboration (2014)\cite{pc}). The CIB map in the GASS field is the cleanest CIB map currently published. We find that with the cross-correlation performed over such a small sky fraction and with the CIB maps containing the dust residuals, we don't detect the ISW signal. In future though, with the availability of the CIB maps with better galactic dust removal, this method will be very powerful to detect the ISW with a high SNR and thereby constraining the dark energy parameters $\Omega_\Lambda$ and $w$. 
\section*{Acknowledgements}
We acknowledge financial support from the \textquotedblleft Programme National de Cosmologie and Galaxies" (PNCG) funded by CNRS/INSU-IN2P3-INP, CEA and CNES, France, from the ANR under the contract ANR-15-CE31-0017 and from the OCEVU Labex (ANR-11-LABX-0060) and the A*MIDEX project (ANR-11-IDEX-0001-02) funded by the \textquotedblleft Investissements d'Avenir" French government programme managed by the ANR.

\end{document}